\input harvmac
\overfullrule=0pt
\Title{\vbox{
\hbox{USC-96/23}
\hbox{hep-th/9610076}
}}
{\vbox{\centerline{Superconformal Fixed Points}
\medskip\centerline{ with $E_n$ Global Symmetry}}}
{\baselineskip=12pt
\centerline{Joseph A. Minahan\foot{minahan@physics.usc.edu} and 
Dennis Nemeschansky\foot{dennisn@physics.usc.edu}}
\bigskip
\centerline{\sl  Department of Physics and Astronomy}
\centerline{\sl University of Southern California}
\centerline{\sl Los Angeles, CA 90089-0484}
\medskip
\bigskip
\centerline{\bf Abstract}

}

\noindent
We obtain the elliptic curve and the Seiberg-Witten differential
 for an $N=2$ superconformal field
theory which has an $E_8$ global symmetry at the strong coupling point
$\tau=e^{\pi i/3}$.  The differential has 120 poles corresponding to
half the charged states in
 the fundamental representation of $E_8$, with the other half living
on the other sheet.  Using this theory, we flow down to $E_7$, $E_6$ and
$D_4$.  A new feature is a $\lambda_{SW}$ for these theories based on
their adjoint representations.  We argue that these theories have
different physics than those with $\lambda_{SW}$ built from the 
fundamental representations.   

\Date{10/96}
\vfil
\eject

\def\NP{{\it Nucl. Phys.\ }}
\def\PL{{\it Phys. Lett.\ }}

\def\PRL{{\it Phys. Rev. Lett.\ }}

\def\vth{\vartheta}

\def\al{\alpha}
\def\be{\beta}
\def\la{\lambda}

\def\lSW{\lambda_{SW}}
\def\ty{\tilde y}
\def\tx{\tilde x}
\def\tb{\tilde\beta}

\def\tT{\widetilde T}
\def\tW{\widetilde W}
\def\hij{h_{ij}}
\def\ij{m_im_j}
\def\thi{{1\over3}}
\def\qu{{1\over4}}
\def\si{{1\over6}}
\def\ei{{1\over8}}
\def\twe{{1\over12}}
\def\trho{\tilde\rho}
\def\hsp{h_{sp}}

\lref\SWI{N. Seiberg and E. Witten, {\it Electric-Magnetic Duality, Monopole 
Condensation, and Confinement in $N=2$ Supersymmetric Yang-Mills Theory},
{\bf hep-th/9407087}, \NP{\bf B426} (1994) {19}.}
\lref\SWII{N. Seiberg and E. Witten, {\it Monopoles, Duality and Chiral
Symmetry Breaking in $N=2$ Supersymmetric QCD}, {\bf hep-th/9408099},
\NP {\bf B431} (1994) {484}.}
\lref\LW{W. Lerche and N. Warner, Exceptional SW Geometry from ALE Fibrations,
 {\bf hep-th/9608183}}
\lref\Vafa{C. Vafa, {\it Evidence for $F$-Theory,} {\bf hep-th/9602022}, 
{\it Nucl. Phys.} {\bf B469} (1996) 403.}
\lref\Sen{A. Sen, {\it $F$-Theory and Orientifolds}, {\bf hep-th/9605150}.}
\lref\KLTY{A. Klemm, W. Lerche, S. Theisen and S. Yankielowicz,
	{\bf hep-th/9411048}, \PL {\bf B344} (1995) {169}.}
\lref\AF{P.C. Argyres and A.E. Faraggi, {\bf hep-th/9411057},
	\PRL {\bf 73} (1995) {3931}.}
\lref\HO{A. Hanany and Y. Oz, {\it On the
	Quantum Moduli Space of Vacua of N=2 Supersymmetric $SU(N_c)$
	Gauge Theories, {\bf hep-th/9505075}, {\it Nucl. Phys.} {\bf B452} 
(1995) 283-312.}}
\lref\APS{P.~C.~Argyres, M.~R.~Plesser and A.~D.~Shapere, 
{\it The Coulomb Phase of N=2 Supersymmetric QCD}, {\bf hep-th/9505100},
{\it Phys. Rev. Lett.} {\bf 75} (1995) 1699-1702.} 
\lref\FK{H.~M.~Farkas and I.~Kra, {\it Riemann Surfaces}, Springer-Verlag 
(1980), New York.}
\lref\Clemens{C.~H.~Clemens, {\it A Scrapbook of Complex Curve Theory}, Plenum
Press (1980), New York.}
\lref\MN{J. Minahan and D. Nemeschansky, 
{\it Hyperelliptic curves for Supersymmetric Yang-Mills}, 
{\bf hep-th/9507032}, \NP{\bf 464} (1996) {3}.}
\lref\MNII{J. Minahan and D. Nemeschansky, 
{\it $N=2$ Super Yang-Mills and Subgroups of $SL(2,Z)$}, 
{\bf hep-th/9601059}, {\it to appear in Nucl. Phys. B}.}
\lref\AD{P. Argyres and M. Douglas, {\it New Phenomena in $SU(3)$ 
Supersymmetric Gauge Theory}, {\bf hep-th/9505062}, 
{\it Nucl. Phys.} {\bf B448} (1995) 93-126}
\lref\AY{O. Aharony and S. Yankielowicz, {\it Exact Electric-Magnetic Duality 
in $N=2$ Supersymmetric QCD Theories}, {\bf hep-th/9601011}.}
\lref\Kob{N. Koblitz, {\it Introduction to Elliptic Curves and Modular Forms},
Springer-Verlag, (1984), New York.}
\lref\Sch{B. Schoeneberg, {\it Elliptic Modular Functions; An Introduction},
Springer-Verlag, (1974), New York.}
\lref\FP{D. Finnell and P. Pouliot, {\it Instanton Calculations Versus Exact 
Results In 4 Dimensional Susy Gauge Theories}, {\bf hep-th/9503115}.}  
\lref\DW{R. Donagi and E. Witten, {\it Supersymmetric Yang-Mills Systems And 
Integrable Systems}, {\bf  hep-th/9510101}.}
\lref\MWI{E. Martinec and N. Warner, {\it Integrable systems and 
supersymmetric gauge theory}, {\bf hep-th/9509161}.}
\lref\Mart{E. Martinec, {\it Integrable Structures in Supersymmetric Gauge 
and String Theory}, {\bf hep-th/9510204}.}
\lref\MWII{E. Martinec and N. Warner, {\it Integrability in N=2 Gauge Theory: 
A Proof}, {\bf hep-th/9511052}.}
\lref\GKMMM{A.Gorsky, I.Krichever, A.Marshakov, A.Mironov and A.Morozov,
{\it Integrability and Seiberg-Witten Exact Solution}, {\bf hep-th/9505035},
{\it Phys. Lett.} {\bf B355} (1995) 466-47.}
\lref\NT{T. Nakatsu and K. Takasaki, {\it Whitham-Toda hierarchy and $N = 2$
 supersymmetric Yang-Mills theory}, {\bf hep-th/9509162}.}
\lref\IM{H. Itoyama and A. Morozov, {\it Integrability and Seiberg-Witten 
Theory: Curves and Period}, {\bf hep-th/9511126}.}
\lref\APSW{P.C. Argyres, M.R. Plesser, N. Seiberg and E. Witten,
{\it New $N=2$ Superconformal Field Theories in Four Dimensions}, 
{\bf  hep-th/9511154}, {\it Nucl.Phys.} {\bf B461} (1996) 71.} 
\lref\EHIY{ T. Eguchi, K. Hori, K. Ito and S.-K. Yang,
{\it Study of $N=2$ Superconformal Field Theories in $4$ Dimensions},
{\bf hep-th/9603002}, {\it Nucl. Phys.} {\bf B471} (1996) 430.}
\lref\DM{K. Dasgupta and S. Mukhi, {\it $F$-Theory at Constant Coupling},
{\bf hep-th/9606044.}}
\lref\Kod{K. Kodaira, {\it On Compact Analytic Surfaces, II-III}, {\it Ann.
of Math.} {\bf 77} (1963) 563; {\bf 78} (1963) 1.}
\lref\MNIII{J.~Minahan and D. Nemeschansky, An N=2 Superconformal Fixed Point with $E_6$ Global Symmetry, {\bf hep-th/9608047.}}
\lref\claude{C. Itzykson, {\it Int. Jour. Mod. Phys. B} {\bf 8} (1994), 3725. }
\lref\KMV{A. Klemm, P. Mayr and C. Vafa, BPS States of Exceptional Non-Critical Strings, {\bf hep-th/9607139.} }
\lref\MNW{J. Minahan, D. Nemeschansky and N. Warner, to appear}
\lref\BDS{T. Banks, M. Douglas and N. Seiberg, Probing F-theory With Branes, 
{\bf hep-th/9605199.} }
\lref\Seib{N. Seiberg, Five Dimensional SUSY Field Theories, Non-trivial Fixed Points and String Dynamics, {\bf hep-th/9608111.} }
\lref\OG{O.~Ganor, Toroidal Compactification of Heterotic 6D Non-Critical Strings Down to Four Dimensions, {\bf hep-th/9608109}.}
\lref\DKV{M. Douglas, S. Katz and C. Vafa, Small Instantons, del Pezzo Surfaces and Type I' theory, {\bf hep-th/9609071.} }
\lref\MS{D. Morrison and N.  Seiberg, Extremal Transitions and Five-Dimensional Supersymmetric Field Theories, {\bf hep-th/9609070.}}
\lref\MV{D. Morrison and C. Vafa, Compactifications of F-Theory on Calabi--Yau Threefolds, {\bf  hep-th/9602114, hep-th/9603161.} }
\lref\Hartshorne{R. Hartshorne, Algebraic Geometry, (Springer, New York, 1974)}

\newsec{Introduction}

Four dimensional $N=2$ superconformal theories with an unbroken $U(1)$
gauge group are  in one to one correspondence with Kodaira's
classification of toroidal singularities.  There are
 7 strong coupling conformal points, which have a global
symmetry that is either $A_0$, $A_1$ or $A_2$, $E_6$, $E_7$ or $E_8$,
or $D_4$.

The $D_4$ case is of course  the celebrated Seiberg-Witten
result for $SU(2)$ Super QCD with four hypermultiplets in the fundamental
representation.  The $A_0$, $A_1$ and $A_2$ cases can be derived from
$D_4$ by taking appropriate limits for the masses in the theory.

The $D_4$ theory has associated with it an elliptic curve and
a differential $\lambda_{SW}$. The deformations of the curve are determined
from the bare masses in the theory as well as the bare coupling.  A
crucial property of $\lambda_{SW}$ is that it has poles in which
the residues are linear combinations of the bare masses.  Furthermore,
it was shown in \SWII\ that this is sufficient to completely
determine the curve.

Of course, the $E_N$ theories cannot be derived from the $D_4$ case.
However, one can assume the existance of $\lambda_{SW}$ for each
of these theories, with the property that the residues of $\lSW$
are linear combinations of $N$ mass parameters that determine the
deformations of the elliptic curve.  In \MNIII, it was shown that
this is enough to completely determine the curve for $E_6$.  In
this paper, we extend this analysis to the cases of $E_8$ and $E_7$.

We will also find an interesting surprise, namely, there exists
other superconformal theories for $E_6$ and $D_4$ as well as $E_7$.  
We find that one can construct a Seiberg Witten differential
based on the adjoints of these groups.  As it happens, the elliptic
curve for the adjoint case is the same as the fundamental.  But since
$\lSW$ is different, the monodromies are different and hence the
content of the physical states are different.  Unlike a gauge symmetry,
a global symmetry is a real symmetry of observable particles, and one
can determine what representation these particles live in.  A $\lSW$
constructed from the adjoint representation will necessarily lead
to physical states living in the adjoint representation.  This also has another
interesting consequence for $D_4$, the adjoint case is invariant under
$SL(2,Z)$, and not just a semi-direct product of $SL(2,Z)$ with
$SO(8)$ triality.

The surfaces that we describe are elliptic fibrations of del Pezzo surfaces.
Such surfaces have appeared in the context of string theory\refs{\MV,\KMV,\LW}
as well as in the study of 5 dimensional gauge theory\refs{\Seib\OG,\DKV,\MS}.
It is hoped that the results presented here will be useful for $F$-theory
considerations\refs{\Vafa,\Sen,\BDS,\DM}, such as the calculations of
BPS masses.

In section 2 we discuss how one finds a set of rational curves for the
case of $E_8$.  In section 3 we discuss the derivation of the $E_8$ curve
and compute $\lSW$.
In section 4 we discuss the flows to the $E_7$, $E_6$ and $D_4$ theories.
In section 5 we present our conclusions.  Most of the results are contained
in the four appendices.

\newsec{Lines, Parabolas and Perfect Squares}

Consider the elliptic curves with an $E_n$ singularity
\eqn\Encurve{\eqalign{
E_6:\qquad\qquad y^2&=x^3-\rho^4\cr
E_7:\qquad\qquad y^2&=x^3-2\rho^3x\cr
E_8:\qquad\qquad y^2&=x^3-2\rho^5,}}
where we have chosen the factors of two for later convenience.  These
curves describe del Pezzo surfaces.
These curves have relevant deformations, the number of which can
be easily found by comparing the dimensions of $x,y$ and $\rho$.
This number is $n$, the rank of the group.  It is convenient to
express these deformations in terms of the $SO(16)$ subgroup for
$E_8$, $SO(12)\times SU(2)$ for $E_7$ and $SO(10)\times U(1)$ for
$E_6$.  Hence, for each of these cases, we have $n$ mass parameters
that live in the Cartan subalgebra of these
subroups.

We also assume that there exists a Seiberg-Witten differential $\lSW$ for
each of these theories, which satisfies
\eqn\sweq{
{d\lSW \over d\rho}\sim {dx\over y}}
and which is allowed to have poles in the $x$ plane
whose residues are linear combinations
of the masses discussed in the previous paragraph.  In order for
this to happen, it must be true that at the positions of the poles,
$y^2$ is a perfect square in terms of $\rho$ and the mass parameters.

In \SWII\ it was shown that such poles can appear in the $D_4$ case
at the positions
\eqn\Dfpole{
x=\beta \rho+\theta}
where $\beta$ is a dimensionless quantity that depends on the bare
coupling and $\theta$ depends on the mass parameters.  The Seiberg-Witten
differential will have four such poles on each sheet, and one is
free to choose a vector, spinor or spinor bar representation for
these poles.

In \MNIII\ it was shown that in the $E_6$ case, the positions of
the poles also satisfy \Dfpole, except in this case, $\beta$ is proportional
to the residue for the particular pole.  There
are 27 such poles, in one to one correspondence with the dimension
of the representation.

As it turns out, the $E_7$ case also has poles described by \Dfpole.
But an inspection of \Encurve\ shows that the poles for the $E_8$
case cannot have this form, since the curve has a $\rho^5$ piece
and hence the leading term in $\rho$ would have an odd power if 
$x$ is linear in $\rho$.  Obviously, $y^2$ cannot be a perfect
square in this situation.  Hence if a Seiberg-Witten differential
is to exist, the poles would have to at least have the form
\eqn\xrhosq{
x=\gamma\rho^2+\beta\rho+\theta,}
in which case $y$ would be cubic in $\rho$.

There is another dilemma involving the poles for $E_8$, and it
is related to the problem of the poles being given by parabolas
and not lines in the $x-\rho$ plane.  The fundamental representation
for $E_8$ is its adjoint.  Accordingly, if some mass parameters
are taken to infinity, the curve should flow to an $E_7$ curve.
The $E_8$ adjoint representation then flows to representations of
the $E_7$ subgroup, which is comprised of the adjoint and two fundamentals.
Hence, if $\lSW$ exists for $E_8$, then there must also exist another
$\lSW$ for $E_7$.  This argument can be extended to $E_6$ and $D_4$,
that is, each of these theories has an $\lSW$ with poles that transform
in the adjoint representation for these groups. In all of these cases
the poles are described by parabolas in the $x-\rho$ plane.

Since the dimension of $\lSW$ is assumed to be one, in all cases, the
dimension of $\rho^2/x$ is two, thus $\gamma$ is dimension negative
two.  As it will turn out $\gamma$ equals $8\pi^2/(Res)^2$, where $Res$ is
the residue for the pole.

We can use homogeneous variables and express the curves in \Encurve\
as curves in the projective space $P^3$. Hence, the rational curves in
\xrhosq\ are of degree 3 in $P^3$.  It is well known that the $E_6$
curve describes a cubic in $P^3$, which is isomorphic to a $P^2$ with
6 points blown up\Hartshorne.  A systematic counting of rational curves has
been carried out for this case (with fixed moduli), where it was found
that there are 72 distinct degree 3 curves (plus another 12 with arithmetic
genus 1)\claude.  These 72 curves transform under the $E_6$ Weyl group.  We
will see that 72 poles do appear when flowing to the $E_6$ case,  with
$y$ a cubic in $\rho$, thus
we see that the 72 poles that we have identified are precisely these curves.
The 12 with arithmetic genus 1 are singlets under the Weyl group, and
so their residues are zero.  

For the $E_7$ and $E_8$ cases, the curves
on the $P^3$ are isomorphic to del Pezzos constructed from $P^2$ with
7 and 8 points blown up.   For the $E_8$ case, the degree 3 curve
in $P^3$ maps
to a curve of degree 1 on the del Pezzo, while for $E_7$, the degree 3 curve
on $P^3$ 
maps to a degree 2 curve.  This must be the case in order to match the counting
of rational curves for these surfaces\refs{\Hartshorne,\KMV}. 

The space of curves with degree higher than 1 have moduli\refs{\claude,\KMV}.
Hence the pole positions for the $E_6$ and $E_7$ adjoints seem to be a
special choice.  The case of generic points in the moduli space is an
interesting question and will be considered elsewhere\MNW.
 

\newsec{The $E_8$ Case}

In this section we explain the derivation of the $E_8$ curve.  All
other theories considered in this paper flow from this one.  Since
$E_8$ has a maximal $SO(16)$ subgroup, we consider deformations
described by eight mass parameters $m_i$.  We define the $SO(16)$ invariants
$T_{2n}$ for $n=1..7$, where
\eqn\Tneq{
T_{2n}=\sum_{0<i_1<i_2..<i_n}^8 m_{i_1}^2..m_{i_n}^2}
There is also the invariant $t_8$,
\eqn\seq{
t_8=\prod_{i=1}^8m_i.}
The $E_8$ curve should be expressible in terms of these independent
$SO(16)$ invariants.

Our method for deriving the curve is to turn on masses one by one,
allowing for all possible terms in the curve consistent with $r$-symmetry, holomorphy and the remaining symmetries.  This still
leaves some ambiguity for the curve.  However, the final terms can
be nailed down by choosing $y^2$ to be a perfect square at the poles,

We choose $x$ such that the curve is of the form
\eqn\gencurve{
y^2=x^3-fx-g.}
By turning on one mass, $m_1$, the symmetry of theory is broken to $SO(14)$.  
Therefore,
we should find a $D_7$ singularity as $\rho$ approaches zero.  Such
a singularity satisfies $g\sim\rho^3$, and $f\sim \rho^2$, and has a 
discriminant $\Delta=4f^3-27g^2\sim\rho^9$.  Up to a rescaling of $m_1$, 
one finds
\eqn\onemass{\eqalign{
f&=m_1^2\rho^3+{m_1^8\over192}\rho^2\cr
g&=2\rho^5+{m_1^6\over24}\rho^4+{m_1^{12}\over 6912}\rho^3.}}

We next assume the ansatz that there are poles at $x=\gamma\rho^2+\beta\rho+
\theta$.  For the curve in \gencurve\ and \onemass, there are two solutions
for $x$ where $y^2$ is a perfect square, 
$x=-\rho^2/m_1^2+\rho m_1^4/12$ and $x=-4\rho^2/m_1^2-\rho m_1^4/24$.
At these points, $y^2$ is given by 
$y^2=-((\rho^3+3\rho^2m_1^6/8)/m_1^3)^2$ and $y^2=-(8\rho^3/m1^3)^2$
respectively.  The form of these solutions is quite suggestive since the
$E_8$ adjoint has as an $SO(16)$ decomposition 
\eqn\repdec{
{\bf 248}={\bf 120}+{\bf 132}.}
Hence, if we assume that $\gamma=-\rho^2/h_{\alpha}^2$, where $h_{\alpha}$ 
is the charge
under the Cartan subalgebra for a particular element of the
representation, then we see that the first pole corresponds
to an $SO(16)$ adjoint and the second pole corresponds to an $SO(16)$
spinor.

Therefore, to ease our search for the $E_8$ curve
we will assume that the contribution of an element of the representation 
to $\lSW$  is proportional to
\eqn\elsw{
{h_{\alpha} y_i\over x-x_{\alpha}}{dx\over y}={\ty_{\alpha}\over h_{\alpha}^2x-\tx_{\alpha}}{dx\over y}}
where $\ty_{\alpha}=h_{\alpha}^3y_i$, and 
\eqn\txeq{
\tx_{\alpha}=h_{\alpha}^2x_{\alpha}=-\rho^2+\tb_i\rho+\tilde \theta_i} 
and the terms $\tx_{\alpha}$ and $\ty_{\alpha}$ are polynomials in the masses and $\rho$.

Turning on another mass $m_2$ breaks the group down to $SO(12)$, in 
which case we should choose the coefficients such that $f\sim \rho^2$,
$g\sim\rho^3$ and $\Delta\sim\rho^8$.   This is not
enough information to determine the terms in $f$ and $g$ and one must
choose the coefficients such that $y^2$ is a perfect square along a 
rational curve in the $x-\rho$ plane.  Assuming that the rational curve
has the form in \txeq\ is sufficient to determine the curve for nonzero
$m_1$ and $m_2$.  In fact, it is enough to only consider the pole with
$h_{\alpha}=m_1+m_2$ to find $f$ and $g$.  With this $f$ and $g$, one then
finds that the other poles corresponding to other elements of the
representation are consistent.

One can keep on turning on masses until the generic deformation is
obtained and the pole positions are determined.  We won't actually
prove here that the results presented below are the unique solutions
to the ansatz in \txeq \foot {The skeptical readers are invited to
download a Mathematica file from http://www.usc.edu/$\sim$minahan /Math/e8.ma and see for themselves that the rational curves lead to perfect squares
in $y^2$ for this deformation of the $E_8$ singular curve.}.

The general idea for computing curves and pole positions is as follows.
If $x$ is chosen to be quadratic in $\rho$, then $y^2$ will be a sextic
equation,
\eqn\ysq{
-y^2=(h_\alpha)^{-6}\sum_{n=0}^6a_n\rho^n,}
where the $a_n$ are polynomials in the $m_i$.  In order that 
$y^2$ be a perfect square, $a_0$ and $a_6$ must be perfect squares of
polynomials involving the $m_i$.  The 
coefficient $a_6$ is always $1$.  The coefficient $a_0$ is
more complicated.  Requiring that it be a perfect square leads to a series
of linear equations for the coefficients of the curve and for the pole
positions.  Once $a_0$ is found, then we look for an $a_1$ such that
the square root of $a_0$ divides $a_1$.  This then leads to more linear
equations for the coefficients.   Finally, we derive more linear equations
by setting to zero the
expression
\eqn\setzero{
\left(\rho^3+\ha a_5\rho^2+\ha {a_1\over\sqrt{a_0}}+\sqrt{a_0}\right)^2
-\sum_{n=0}^6a_n\rho^n.}
This turns out to be sufficient for determining the complete curve and
the poles.

The final results for the $E_8$ curve are presented in the appendix A.  It
is convenient to express the curve in terms of a different $SO(16)$ invariant
$\tT_4=T_2^2/4-T_4$.  This gets rid of most of the higher powers in $T_2$.
By inspection, one sees that most of the generic terms in the curve
actually have zero coefficient, which is a good thing, otherwise the
expression for $g$ alone would have 341 terms instead of the much
more manageable 71 terms.

One still has the freedom to shift $\rho$, removing the $\rho^4$
term in $g$.  After this shift, 
$\rho\to \rho-(T_2\tT_4/6+T_6)$, the coefficients of $\rho$ in
$f$ and $g$ we are left with
are eight independent Casimirs of $E_8$ and hence form
a natural basis for the entire set of $E_8$ casimirs.

We have also given the positions of the poles as well as the corresponding
values for $y$  in the appendix.  For an
$SO(16)$ adjoint pole, the pole position should be expressible in
terms of two masses, $m_i$ and $m_j$, and the casimirs for the
unbroken $SO(12)$ orthogonal to $i$ and $j$, $W_n$ and $W_6$.  

The residue of
the spinor poles are given by 
${1\over2}{1\over2\sqrt{2}\pi i}\sum_i \pm m_i$, where
the number of $-$ signs is even.  The state with all $+$ signs
has a residue that is proportional to the linear symmetric polynomial of
the eight masses.  Hence this pole position is expressible in terms
of the symmetric polynomials $s_n$, where
\eqn\sympol{
s_n=\sum_{0<i_1..<i_n} m_{i_1}...m_{i_n}
}
In order to show that this pole leads to a perfect square,
we need the relation
\eqn\Tsrel{
T_{2n}=s_n^2+2\sum_{m=1}^n (-1)^m s_{n-m}s_{n+m},}
where $s_0=1$ and $s_p=0$ if $p>8$.
The poles for the rest of the spinor states can be found by changing
an even number of the signs of $m_i$ in the symmetric polynomials.  Because
the curve has $t_8$ dependence, there are no poles at
the positions found by changing an odd number of the signs.  Hence,
only one spinor of $SO(16)$ appears, as is expected.

There are some checks that we can do for our curve.  We can let all the
masses satisfy $m_i=m$ and compute $f$ and $g$.  In
this case, we find that 
\eqn\eqmass{
f\sim (\rho+4m^6)^3\qquad\qquad g\sim (\rho+4m^6)^5.}
This is the behavior for an $E_7$ singularity.  A simple counting shows
that this is sensible.  The states with residues proportional to $m_i-m_j$ are
massless, as are the spinor states with an equal number of plus and
minus signs.  There are 56 states of the former type and 70 of the
latter, leaving 126 massless states, the number of charged states
in the $E_7$ adjoint.  Likewise, we can consider the case where $m_i=m,\ i<8$
and $m_8=-m$.  In this case there is an $A_7$ singularity.
There are still 56 massless states coming from the adjoint,
but now, none of the spinor states are massless since an odd number of $-$
signs is not allowed.  Hence the counting is consistent with the number
of charged states in the adjoint of $SO(8)$.

Once we have the pole positions and the value of $y^2$ at these poles
we can sum these contributions to the Seiberg-Witten differential.  
As we have already mentioned, if the masses are such that a residue
is zero, because of the form of the sum, the corresponding term 
can still contribute to $\lSW$.  However, an interesting
feature occurs in this situation.  By inspection of eqs. (A.3-A.7),
one sees that if the residue is zero then $\tx_{\alpha}$ divides
$\ty_{\alpha}$, leaving a term that is linear in $\rho$ in front of
${dx/ y}$.  Furthermore, the coefficient of $\rho$ is the same for
any state.  It is also true that only the charged states are summed
over, so it is unlikely that the sum of the poles is the complete
Seiberg-Witten differential.  However, the new piece should be 
invariant under the $E_8$ Weyl group and should be at most linear in
$\rho$.

Let us thus assume that $\lSW$ is given by
\eqn\sweqpre{
\lSW={1\over2\sqrt{2}\pi}(A\rho+BT_2^3+CT_2T_4+DT_6){dx\over y}
+{1\over2\sqrt{2}\pi i}
\sum_{\al=1}^{120}{\ty_\al\over h_\al^2x-\tx_\al}{dx\over y}
}
where the sum is over half of the 240 charged states of the representation
and $h_\al/(2\sqrt{2}\pi i)$ is the residue of the state, 
normalized such that an
$SO(16)$ adjoint state has $h_\al=\pm m_i\pm m_j$ and the
spinor state has $h_\al=\sum_i\pm m_i/2$, where the number of $-$
signs is even.  The factor
in front of the sum is chosen to match the normalization in \SWII.
The coefficients $A$, $B$ and $C$ are found by fixing 
\eqn\swder{
{d\lSW\over d\rho}= {k\over\sqrt{2}\pi}{ dx\over y},
}
up to a total derivative, where $k$ is to be determined.  
Given this form the coefficients in $\lSW$
can be derived by considering special values for the $m_i$.  In
particular, letting $m_1=m, m_i=0, i>1$ determines $A$ and $k$, 
$m_1=m_2=m, m_i=0\ i>2$ determines $B$ and $m_1=m_2=m_3=m$ determines $C$.
We find that $k=30$, the Coxeter number for $E_8$ (half the index of
the adjoint representation), and that $\lSW$ is
\eqn\sweq{
\lSW={1\over2\sqrt{2}\pi}(60\rho-T_2T_4+6T_6){dx\over y}
+{1\over2\sqrt{2}\pi i}
\sum_{\al=1}^{120}{\ty_\al\over h_\al^2x-\tx_\al}{dx\over y}.
}
Note that the piece without the poles is 
$${1\over2\sqrt{2}\pi}\left(60\tilde\rho-{1\over4}T_2^3\right){dx\over y}$$
where $\tilde\rho$ is the value of $\rho$ after shifting to remove
the $\rho^4$ term in $g$.  $T_2$ is invariant under the $E_8$ Weyl group,
therefore, the entire term is invariant.

\newsec{Flowing to the Other Cases}

We can investigate the $E_7$ theory by taking two of the $E_8$ masses
to infinity.  Accordingly, let $m_7=\Lambda-\phi/2$ and $m_8=\Lambda+\phi/2$.
These variables are the natural variables for the $E_7$ subgroup $SO(12)\times
SU(2)$.
We also rescale $x$, $y$ and $\rho$ by $x\to \Lambda^4\rho$, $y\to\Lambda^6 y$
and $\rho\to \Lambda^2\rho$.  Plugging these new values into (A.1) and (A.2)
and keeping the terms to leading order in $\Lambda$ gives the $E_7$
curve.  The values of $f$ and $g$ are given in (B.1) and (B.2),where
 now the terms in the
curve are given in terms of $D_6$ invariants.  It is convenient to 
replace $T_2$ by $\tT_2=T_2-\phi^2$.  In order to express the curve
explicitly in terms of $E_7$ invariants, it is necessary to shift $\rho$
by $(\tT_2)^2/72+T_4/6$, which removes the $\rho^2$ term in $f_{E_7}$.

The positions of the poles are also found by using these same scaling
arguments.  However, since the adjoint of $E_8$ decomposes under its
$E_7\times SU(2)$ subgroup to
\eqn\adjoint{
{\bf 248}={\bf(133,1)}+{\bf(1,3)}+{\bf(56,2)}
}
some of these poles will flow to $E_7$ adjoints and others will flow
to $E_7$ fundamentals.  In fact, it is easy to check that for the
fundamentals, the $\rho^2$ piece scales out of $x_\al$, leaving a 
linear relation between $x$ and $\rho$.  

It should be possible to find a self-consistent $\lSW$ for each representation.  Clearly, the coefficient $k$ in \sweqpre\ should split
into a contribution of one $E_7$ adjoint and two fundamentals.  Since
$k$ is the Coxeter number, this picture is consistent with $k$ splitting
into 18, the $E_7$ coxeter number,  and two values of 6, half the index
 of the fundamental representation.

For the adjoint case, the $\lSW$ is of the form
\eqn\sweqpresev{
\lSW={1\over2\sqrt{2}\pi}(A\rho+BT_2^2+C\phi^2T_2+D\phi^4+ET_4){dx\over y}
+{1\over2\sqrt{2}\pi i}
\sum_{\al=1}^{63}{\ty_\al\over h_\al^2x-\tx_\al}{dx\over y}
}
where the sum is over half of the 126 charged states of the representation.
The values for $\tx_\al$ and $\ty_\al$ are found in the appendix.  
In terms of the $SO(12)\times
SU(2)$ subgroup, the residues of the poles are of the form 
$(\pm m_i\pm m_j)/(2\sqrt{2}\pi i)$,
$\pm \phi/(2\sqrt{2}\pi i)$ or $\sum_j( \pm m_j\pm\phi/2)/(2\sqrt{2}\pi i)$, 
with an odd number of $-$ signs
in front of the $m_i$ for the $SO(12)$ spinor.
As in the $E_8$ case, the coefficients $A$,$B$, $C$, $D$, and $E$ are chosen
so that $d\lSW/d\rho\sim dx/y$.  Again, one can find these coefficients
by choosing special values for the masses.  The final result in
this case is that
\eqn\sweqsev{
\lSW={1\over2\sqrt{2}\pi}\left(9\rho+{1\over8}(\tT_2)^2+{3\over2}T_4\right)
{dx\over y}+{1\over2\sqrt{2}\pi i}
\sum_{\al=1}^{63}{\ty_\al\over h_\al^2x-\tx_\al}{dx\over y},
}
and with $k=18$, the Coxeter number for $E_7$.  The residues for these
poles are $(\pm m_i\pm\phi/2)/(2\sqrt{2}\pi i)$ or 
$\sum_j \pm m_i/(4\sqrt{2}\pi i)$ with an even number of
$-$ signs for the spinors.
The term without the poles is clearly proportional to the shifted value
of $\rho$, and hence this term is clearly an $E_7$ invariant. 

For the fundamental case, the positions of the poles are now linear in
$\rho$ since  the quadratic piece scaled out for these particular states. 
One can do an analysis similar to the adjoint case, with the result
\eqn\sweqsevf{
\lSW={1\over2\sqrt{2}\pi}\left(24\rho+{1\over3}(\tT_2)^2+{3\over2}T_4
-3\left({2\over3}T_2+\phi\right)^2\right)
{dx\over y}+{1\over2\sqrt{2}\pi i}
\sum_{\al=1}^{28}{h_\al y_\al\over x-x_\al}{dx\over y},
}
where the sum is over half the 56 states.  The value of $k$ is found to
be $k=6$. 
The term without poles in \sweqsevf\ is comprised of a shifted $\rho$,
plus a piece that is proportional to the square of the $E_7$ casimir
of weight 2, hence the entire term is an $E_7$ invariant.

 A natural way to
flow from the $E_7$ theory to the $E_6$ curve is under the scaling
 $m_6=\Lambda$, $\phi=\Lambda-6\lambda$, $x\to x\Lambda^2$, $y\to y\Lambda^3$,
and $\rho\to \rho\Lambda$.  The masses $m_i$ and $\lambda$ are the
natural variables of the 
$SO(10)\times U(1)$ subgroup of $E_6$.  Keeping only the leading orders
in $\Lambda$ gives the values of $f$ and $g$ in (C.1) and (C.2).

For the adjoint case, the Seiberg Witten differential is given by
\eqn\sweqsix{
\lSW={1\over2\sqrt{2}\pi i}
\sum_{\al=1}^{36}{\ty_\al\over h_\al^2x-\tx_\al}{dx\over y},
}
where the sum is over half of the charged states in the representation.
The value for $k$ is found to be $k=12$, the $E_6$ Coxeter number.  The 
residues of these states are $(\pm m_i\pm m_j)/(2\sqrt{2}\pi i)$, and the 
spinors
 $\sum_j (\pm m_i/2\pm3\lambda)/(2\sqrt{2}\pi i)$,
with an odd number of $-$ signs for $+3\lambda$ and an even number otherwise.
Curiously, the sum over the poles is the complete differential.  Unlike
the $E_8$ and $E_7$ case, there is no extra piece linear in $\rho$.

The $E_6$ fundamental $\lSW$ was given in $\MNIII$ and was found to
be
\eqn\sweqsixf{
\lSW={1\over2\sqrt{2}\pi}\left(18\rho+18\lambda T_2-72\lambda^3\right)
{dx\over y}+{1\over2\sqrt{2}\pi i}
\sum_{\al=1}^{27}{h_\al y_\al\over x-x_\al}{dx\over y},
}
where the sum is over the 27 charged states of the $E_6$ fundamental.
Since this is a complex representation, the residues on the other sheet
are part of the conjugate representation.  Because of this, $k=6$,
which is the index for the fundamental representation instead of half
the index.  The residues are of the form $+4\lambda/(2\sqrt{2}\pi i)$, 
$(\pm m_i -2\lambda)/(2\sqrt{2}\pi i)$
or $\sum_j(\pm m_j/2 +\lambda)/(2\sqrt{2}\pi i)$, 
where the number of $-$ signs is even.

Finally, we come to the $D_4$ case.  This flow was discussed in \MNIII.
The appropriate scaling is to let $x\to x\Lambda^2$, $y\to y\Lambda^3$
and to set $\rho=u\Lambda$, $\la=c_1\Lambda/6$ and $m_5=-c_2\Lambda$,
where $c_1$ and $c_2$ are the combination of theta functions defined
in \SWII,
\eqn\ceq{
c_1=\ha\left(\vth_3^4(\tau)+\vth_2^4(\tau)\right)\qquad\qquad
c_2=\ha\vth_1^4(\tau).}
Keeping only the leading powers in $\Lambda$, $f$ and $g$ reduce to
the expressions in (D.1) and (D.2).  The adjoint pole positions and
values for $y$ at the poles are given in (D.3) and (D.4). 

The Seiberg Witten differential for the adjoint representation is
similar in form to the $E_6$ expression, with $\lSW$ given by
\eqn\sweqfour{
\lSW={1\over2\sqrt{2}\pi i}
\sum_{\al=1}^{12}{\ty_\al\over h_\al^2x-\tx_\al}{dx\over y},
}
where the sum is over half of the 24 charged states in the representation.
We also find that $k=6$, the $SO(8)$ coxeter number.  As in the case
for $E_6$, the sum over poles is the entire $\lSW$, there is no extra
holomorphic piece.

\newsec{Discussion}

In this paper we have constructed superconformal theories with $E_n$
global symmetries.  We have also constructed Seiberg-Witten differentials
based on the adjoints of these groups, as well as $\lSW$ for an $SO(8)$
global symmetry.  

The natural question arises whether or not the theories
are equivalent to theories where the Seiberg-Witten differential is
constructed from the fundamental representation of these groups.  At first
one might think that they are equivalent since the elliptic curves are
the same.  So one immediately concludes that the coupling is the same
if all parameters are the same.  However, the masses of the BPS states
are found from $\lSW$ and here it seems that there could be differences.
For instance, consider the $SO(8)$ case with all $m_i$ set to $0$.
Then $\lSW$ looks identical for the vector, spinor or adjoint rep.
However, the adjoint $\lSW$ has a different normalization, so it would
seem that all BPS states that one finds are 6 times heavier than those
in the theory with a  vector $\lSW$, since $k=6$ for the adjoint and $k=1$
for the vector.

If we tried to divide by this factor of 6 to set the masses equal, then
another problem arises when we turn on the $m_i$.  Then we find
that there are monodromies such that the coordinates $a$ or $a_D$
shift by $(\pm m_i\pm m_j)/(6\sqrt{2})$.  No such shifts are possible
for the vector $\lSW$.  So we must conclude that the theories are different.

Still, the behavior is surprising for $D_4$.  The standard lore is that
for each pole in $\lSW$, there is an electric state with charge $1$.
Hence the electric states are transforming under the adjoint of $SO(8)$.
But by triality, the magnetic and dyonic states
also transform under the adjoint representation.  We still find that
the electric coupling runs to zero in the same fashion as in \SWII,
even though the electric states are different.  The resolution of
this paradox must be that the monopoles and dyons somehow contribute
to the $\be$-function.  

{\bf Acknowledgements}: 
We thank Nick Warner for many helpful conversations.
This research was supported in part by D.O.E.~grant DE-FG03-84ER-40168.

\vfill\eject
 
\appendix{A}{$E_8$ results}

For the curve of the form $y^2=x^3-fx-g$, the curve with $E_8$ symmetry
has $f$ and $g$ given by
\eqn\feq{\eqalign{f=&
 {\rho^3}\,{  T_2}+{\rho^2}\,\left( 14\,{  t_8} + 
     {{{{{  (\tT_4)}}^2}}\over {12}} + {  T_8} \right)+
  \rho
\,\left( 8\,{  T_{14}} - {  T_{12}}\,{  T_2} - 
     {{{ 2 T_{10}}\,{  \tT_4}}\over 3} + 
     {{11\,{  t_8}\,{  T_2}\,{  \tT_4}}\over {3}} + 
     8\,{  t_8}\,{  T_6} \right) \cr&
 - {  T_{12}}\,{  T_8} - 
  {{  t_8}\,{  \tT_4}\,{  T_8}} +
{{{{{  T_{10}}}^2}}\over 3} + 2\,{  t_8}\,{  T_{12}} - 
{{2\,{  t_8}\,{  T_{10}}\,{  T_2}}\over 
    3} + {{{{{  t_8}}^2}\,{{{  T_2}}^2}}\over 3} \cr& 
 + {{{{2  t_8}}^2}\,{  \tT_4}} + 
  {{2  T_{14}}\,{  T_2}\,{  \tT_4}} + 
  {{{  T_{12}}\,{{{  (\tT_4)}}^2}}\over {4}} + 
  {{{  t_8}\,{{{  (\tT_4)}}^3}}\over {4}} + 
  4\,{  T_{14}}\,{  T_6}
  }
}
\eqn\geq{\eqalign{
g&=
2\,{\rho^5} + 
  {\rho^4}\,\left( {{{  T_2}\,{  \tT_4}}\over {6}} + 
     {  T_6} \right)
 + 
  {\rho^3}\,\left( -4\,{  T_{12}} + 
     {{{  T_{10}}\,{  T_2}}\over 3} + 
     {{5\,{  t_8}\,{{{  T_2}}^2}}\over 3} + 
     {{22\,{  t_8}\,{  \tT_4}}\over 3} + 
     {{{{{  (\tT_4)}}^3}}\over {108}} - 
     {{{  \tT_4}\,{  T_8}}\over {3}} \right)\cr&
 + 
  {\rho^2}\,\biggl( {{-34\,{  t_8}\,{  T_{10}}}\over 3} + 
     {{10\,{{{  t_8}}^2}\,{  T_2}}\over 3} + 
     {  T_{14}}\,{{{  T_2}}^2} + 
     {{16\,{  T_{14}}\,{  \tT_4}}\over 3} - 
     {{{ 2 T_{12}}\,{  T_2}\,{  \tT_4}}\over 3} + 
     {{5\,{  T_{10}}\,{{{  (\tT_4)}}^2}}\over {36}} + 
     {{31\,{  t_8}\,{  T_2}\,{{{  (\tT_4)}}^2}}\over {36}}\cr&
 -    2\,{  T_{12}}\,{  T_6} + 
     {{10\,{  t_8}\,{  \tT_4}\,{  T_6}}\over 3} + 
     {{{  T_{10}}\,{  T_8}}\over 3} + 
     {{11\,{  t_8}\,{  T_2}\,{  T_8}}\over 3} \biggr) \cr&
 + 
  \rho\,\biggl( 8\,{{{  t_8}}^3} + 2\,{{{  T_{12}}}^2} - 
     {{16\,{  T_{10}}\,{  T_{14}}}\over 3} - 
     {{{  T_{10}}\,{  T_{12}}\,{  T_2}}\over 3} + 
     {{4\,{  t_8}\,{  T_{14}}\,{  T_2}}\over 3} + 
     {{{  t_8}\,{  T_{12}}\,{{{  T_2}}^2}}\over 3} - 
     {{{{{ 2 T_{10}}}^2}\,{  \tT_4}}\over {9}}\cr& + 
     {{10\,{  t_8}\,{  T_{12}}\,{  \tT_4}}\over 3} - 
     {{23\,{  t_8}\,{  T_{10}}\,{  T_2}\,{  \tT_4}}\over 
       {9}} + {{25\,{{{  t_8}}^2}\,{{{  T_2}}^2}\,
	 {  \tT_4}}\over {9}} + 
     {{10\,{{{  t_8}}^2}\,{{{  (\tT_4)}}^2}}\over {3}} + 
     {{5\,{  T_{14}}\,{  T_2}\,{{{  (\tT_4)}}^2}}\over {6}} - 
     {{{  T_{12}}\,{{{  (\tT_4)}}^3}}\over {12}}\cr& + 
     {{{  t_8}\,{{{  (\tT_4)}}^4}}\over {24}} - 
     {{16\,{  t_8}\,{  T_{10}}\,{  T_6}}\over 3} + 
     {{16\,{{{  t_8}}^2}\,{  T_2}\,{  T_6}}\over 3} + 
     {{8\,{  T_{14}}\,{  \tT_4}\,{  T_6}}\over 3} - 
     8\,{{{  t_8}}^2}\,{  T_8} + 
     2\,{  T_{14}}\,{  T_2}\,{  T_8}\cr& + 
     {{{  T_{12}}\,{  \tT_4}\,{  T_8}}\over {3}} - 
     {{{ 2 t_8}\,{{{  (\tT_4)}}^2}\,{  T_8}}\over {3}} + 
     2\,{  t_8}\,{{{  T_8}}^2} \biggr)\cr&
 + {{2\,{{{  T_{10}}}^3}}\over {27}} + 
  {{2\,{  t_8}\,{  T_{10}}\,{  T_{12}}}\over 3} + 
  4\,{{{  t_8}}^2}\,{  T_{14}} - 
  {{2\,{  t_8}\,{{{  T_{10}}}^2}\,{  T_2}}\over 9} - 
  {{2\,{{{  t_8}}^2}\,{  T_{12}}\,{  T_2}}\over 3} + 
  {{2\,{{{  t_8}}^2}\,{  T_{10}}\,{{{  T_2}}^2}}\over 9}
 -   {{2\,{{{  t_8}}^3}\,{{{  T_2}}^3}}\over {27}} \cr&
+  {{{{{ 2 t_8}}^2}\,{  T_{10}}\,{  \tT_4}}\over 3} - 
  {{{{{2  t_8}}^3}\,{  T_2}\,{  \tT_4}}\over 3} + 
  {{{{{  T_{12}}}^2}\,{  T_2}\,{  \tT_4}}\over 2} - 
  {{{4  T_{10}}\,{  T_{14}}\,{  T_2}\,{  \tT_4}}\over 3} + 
  {{{4  t_8}\,{  T_{14}}\,{{{  T_2}}^2}\,{  \tT_4}}\over 3} + 
  {{{  T_{10}}\,{  T_{12}}\,{{{  (\tT_4)}}^2}}\over {12}}\cr&
 + {{  t_8}\,{  T_{14}}\,{{{  (\tT_4)}}^2}} + 
  {{11\,{  t_8}\,{  T_{12}}\,{  T_2}\,{{{  (\tT_4)}}^2}}\over 
    {12}} + {{{  t_8}\,{  T_{10}}\,{{{  (\tT_4)}}^3}}\over 
    {12}} + {{5\,{{{  t_8}}^2}\,{  T_2}\,
      {{{  (\tT_4)}}^3}}\over {12}} + 
  {{{  T_{14}}\,{{{  (\tT_4)}}^4}}\over {16}} + 
  {{{  T_{12}}}^2}\,{  T_6}\cr&
 -   {{8\,{  T_{10}}\,{  T_{14}}\,{  T_6}}\over 3} + 
  {{8\,{  t_8}\,{  T_{14}}\,{  T_2}\,{  T_6}}\over 3}
 +   {{2  t_8}\,{  T_{12}}\,{  \tT_4}\,{  T_6}} + 
  {{{{  t_8}}^2}\,{{{  (\tT_4)}}^2}\,{  T_6}}  - 
  {{{  T_{10}}\,{  T_{12}}\,{  T_8}}\over 3}\cr& - 
  4\,{  t_8}\,{  T_{14}}\,{  T_8} + 
  {{{  t_8}\,{  T_{12}}\,{  T_2}\,{  T_8}}\over 3}
 -   {{{  t_8}\,{  T_{10}}\,{  \tT_4}\,{  T_8}}\over {3}} + 
  {{{{{  t_8}}^2}\,{  T_2}\,{  \tT_4}\,{  T_8}}\over 
    {3}} - {{{  T_{14}}\,{{{  (\tT_4)}}^2}\,{  T_8}}\over 
    {2}} + {  T_{14}}\,{{{  T_8}}^2} }
}
where the $SO(16)$ invariants $T_n$, $t_8$ and $\tT_4$ are described
in the text.

The 240 poles occur at the 112 charged states in the adjoint of $SO(16)$
and the 128 states of a spinor representation.  In finding the pole
positions, it is convenient to expand the expression in powers of
$h_\al$, where $h_\al/(2\sqrt{2}\pi i)$ is the residue for that pole.  
For the adjoint case
we find that $\tx_{ij}=\hij^2x_{ij}$ satisfies
\eqn\xadj{\eqalign{
 \tx_{ij}=&  -\left(\rho+\ha\ij\tW_4-w_6\right)^2 
 +\hij^2\biggl(\thi\rho\tW_4-w_6\tW_4 +{2\over3}W_{10}\cr&
 \qquad+\ij\left({2\over3}W_8-{1\over6}\rho W_2+{5\over6}W_2w_6\right)
 +(\ij)^2\left({4\over3}w_6+{2\over 3}W_6 +\qu W_2\tW_4\right)\biggr)\cr
 &-\hij^4\left(\ha W_2w_6+\si\rho W_2+\thi W_8+{1\over12}\ij(29w_6-5\rho)
 +{1\over16}(\ij)^2(W_2^2+2\tW_4)\right)\cr&
 +\hij^6\left({3\over4}w_6+\twe\rho
 +{1\over16}(\ij)^2W_2\right)-{1\over64}\hij^8(\ij)^2
 }}
where $h_{ij}=(m_i+m_j)$ and the variables $W_n$ satisfy    
\eqn\Weqs{
W_n=\sum_{i_1<..i_n\atop i\ne i_k\ne j}m^2_{i_1}..m^2_{i_n}\qquad\qquad
\tW_4=\qu W_2^2-W_4\qquad\qquad w_6=t_8/(\ij)
}
We could also choose to change the sign of $m_j$ in these expressions.
When $x=x_{ij}$ then $y^2$ is a perfect square, with $\sqrt{-y_{ij}^2}$ 
given by
\vfill\eject
\eqn\yadj{\eqalign{
&\sqrt{-y^2_{ij}}=\cr
&(\rho-w_6+\ha\ij\tW_4)^3-\ei\hij^2\left(\rho-w_6+\ha\ij\tW_4\right)
\times\cr
&\ \times\biggl(4\rho\tW_4+8W_{10}
-12w_6\tW_4+2\ij(4W_8+5W_2w_6-\rho W_2)+(\ij)^2(3W_2\tW_4+8W_6+16w_6)
\biggr)\cr
&+{1\over32}\hij^4\biggl(16W_{10}\tW_4-8w_6(\tW_4)^2
-8\rho^2W_2-32w_6W_8-24W_2w_6^2\cr
&\qquad\qquad
+\ij(12\rho^2+8\rho W_2\tW_4+32\rho W_6-56\rho w_6-84w_6^2-24W_2\tW_4w_6-
24W_2W_{10}-96w_6W_6\cr
&\qquad\qquad+(\ij)^2(-4\tW_4\rho-2W_2^2\rho+20\tW_4w_6-8W_2W_8
-64W_{10}-6W_2^2w_6)\cr
&\qquad\qquad+(\ij)^3(3W_2^2\tW_4-32W_8+8W_2W_6+3(\tW_4)^2)\biggr)\cr
&+{1\over8}\hij^6\biggl(3\rho^2+12\rho w_6+9w_6^2+2W_2\tW_4w_6+8w_6W_6
+2W_2W_{10}\cr
&
+{1\over2}\ij(-\rho W_2^2-2\rho\tW_4+3W_2^2w_6+6\tW_4w_6+30W_{10})
+\ha(\ij)^2(3\rho W_2+7W_2w_6-6W_8)\cr
&\qquad\qquad+\ei(\ij)^3(48w_6-8W_6-W_2^3-6W_2\tW_4)
\biggr)\cr
&+{1\over128}\hij^8\biggl(-48W_{10}-8W_2^2w_6-16\tW_4w_6+8\ij(\rho W_2-3W_2w_6)
-2(\ij)^2(5\rho+19w_6)\cr
&\qquad\qquad+3(\ij)^3(W_2^2+\tW_4)\biggr)\cr
&
+{1\over256}\hij^{10}(16W_2w_6-4\ij(\rho-3w_6)-3(\ij)^3W_2)
-{1\over512}\hij^{12}(8w_6-(\ij)^3)
}}

The spinor poles are found at $\tx_{sp}=h_{sp}^2x_{sp}$, where $h_{sp}=s_1/2$
and
\eqn\xspin{\eqalign{
\tx_{sp}&=-\trho^2-s_1s_5\trho+\twe s_1^2(4\trho s_4-s_5^2-8s_3s_7)
+\twe s_1^3(-\trho s_3+2s_3s_6+6s_2s_7)\cr
&+{1\over24}s_1^4(\trho s_2-2s_2s_6-4s_8)
-\ei s_1^5s_7-{1\over96}s_1^6(\trho-2s_6),
}}
where $s_n$ is the symmetric polynomial for eight masses of order $n$, 
linear in each $m_i$, and with
an even number of $m_i$ replaced with $-m_i$. The term $\trho$
is $\trho=\rho-s_6$.  At the poles, $y^2$ is a perfect square, with
$\ty_{sp}=y_{sp}h_{sp}^3$ satisfying
\vfill\eject
\eqn\yspin{\eqalign{
\sqrt{-\ty_{sp}^2}=&\trho^3+{2\over3}s_1s_5\trho^2-\ha s_1^2\trho
(\trho s_4-2s_3s_7-s_5^2)\cr&
+\ei s_1^3(\trho^2s_3-2\trho( s_3s_6+3s_2s_7+s_4s_5)+4s_3s_5s_7)\cr
&+\ei s_1^4(\trho(s_3s_5+2s_2s_6-2s_8)-2s_3s_4s_7-2s_2s_5s_7-2s_3^2s_8-2s_7^2)
\cr
&+{1\over16} s_1^5(\trho(4s_7-s_2s_5)+6s_2s_3s_8+2s_2s_4s_7+2s_6s_7-2s_5s_8
+2s_3^2s_7)\cr
&+{1\over16}s_1^6(-\trho s_6-2s_2^2s_8+s_5s_7-2s_2s_3s_7)
+{1\over64}s_1^7(\trho s_5+2s_2^2s_7-2s_4s_7-6s_3s_8)\cr
&+{1\over32}s_1^8(2s_2s_8+s_3s_7)-{1\over64}s_1^9s_2s_7
-{1\over128}s_1^{10}s_8+{1\over256}s_1^{11}s_7
}}
Notice that in \xadj-\yspin, when $h_\al$ is zero, then
$\tx_\al$ divides $\ty_\al$.

\appendix{B}{$E_7$ Results}

 The $E_7$ curve is derived from the $E_8$ curve by setting 
 $m_7=\Lambda-\phi/2$ and $m_8=\Lambda+\phi/2$ and then taking the
 limit $\Lambda\to\infty$.  Scaling the other variables as described
 in the text, and keeping only the leading order terms, one finds
\eqn\feqEsev{\eqalign{
f_{E_7}&=2\,{\rho^3} + 
  {\rho^2}\,\left( {{{{{ (\tT_2)}}^2}}\over {12}} + { T_4}
      \right) + 
  \rho\,\left( 8\,{\phi^2}\,{ t_6} + 
     {{2\,{ \tT_2}\,{ t_6}}\over 3} + 
     {{2\,{ \tT_2}\,{ T_6}}\over 3} - 2\,{ T_8} \right) \cr&
 + 4\,{\phi^2}\,{ T_{10}}  - {{{{{ (\tT_2)}}^3}\,{ t_6}}\over 4} + 
  { \tT_2}\,{ T_4}\,{ t_6} + 
  {{4\,{{{ t_6}}^2}}\over 3} - 
  {{4\,{ t_6}\,{ T_6}}\over 3} + 
  {{{{{ T_6}}^2}}\over 3} 
   + {{{{{ (\tT_2)}}^2}\,{ T_8}}\over 4} - { T_4}\,{ T_8}
}
}
and
\vfill\eject
\eqn\geqEsev{\eqalign{
g_{E_7}=&{\rho^4}\,\left( {\phi^2} + {{2\,{ \tT_2}}\over 3} \right) + 
  {\rho^3}\,\left( {{-{{{ (\tT_2)}}^3}}\over {108}} + 
     {{{ \tT_2}\,{ T_4}}\over 3} + 
     {{20\,{ t_6}}\over 3} + {{2\,{ T_6}}\over 3} \right) \cr& + 
  {\rho^2}\,\biggl( 4\,{ T_{10}} - 
     {{10\,{\phi^2}\,{ \tT_2}\,{ t_6}}\over 3} - 
     {{29\,{{{ (\tT_2)}}^2}\,{ t_6}}\over {18}} + 
     {{22\,{ T_4}\,{ t_6}}\over 3} + 
     {{5\,{{{ (\tT_2)}}^2}\,{ T_6}}\over {36}} + 
     {{{ T_4}\,{ T_6}}\over 3}\cr& - 2\,{\phi^2}\,{ T_8} - 
     {{2\,{ \tT_2}\,{ T_8}}\over 3} \biggr) \cr&  + 
  \rho\,\biggl( {{-8\,{\phi^2}\,{ \tT_2}\,{ T_{10}}}\over 3} - 
     {{{ (\tT_2)}}^2}\,{ T_{10}} + 4\,{ T_{10}}\,{ T_4} + 
     {{{{{ (\tT_2)}}^4}\,{ t_6}}\over {24}} - 
     {{2\,{{{ (\tT_2)}}^2}\,{ T_4}\,{ t_6}}\over 3}\cr& + 
     2\,{{{ T_4}}^2}\,{ t_6} + 
     {{32\,{\phi^2}\,{{{ t_6}}^2}}\over 3} - 
     {{4\,{ \tT_2}\,{{{ t_6}}^2}}\over 9} - 
     {{16\,{\phi^2}\,{ t_6}\,{ T_6}}\over 3} - 
     {{2\,{ \tT_2}\,{ t_6}\,{ T_6}}\over 9}\cr& + 
     {{2\,{ \tT_2}\,{{{ T_6}}^2}}\over 9} + 
     {{{{{ (\tT_2)}}^3}\,{ T_8}}\over {12}} - 
     {{{ \tT_2}\,{ T_4}\,{ T_8}}\over 3} + 
     {{4\,{ t_6}\,{ T_8}}\over 3} - 
     {{2\,{ T_6}\,{ T_8}}\over 3} \biggr)\cr& + 
  {{{{{ (\tT_2)}}^4}\,{ T_{10}}}\over {16}} - 
  {{{{{ (\tT_2)}}^2}\,{ T_{10}}\,{ T_4}}\over 2} + 
  { T_{10}}\,{{{ T_4}}^2} + 
  {{16\,{\phi^2}\,{ T_{10}}\,{ t_6}}\over 3} + 
  {\phi^2}\,{{{ (\tT_2)}}^2}\,{{{ t_6}}^2}\cr& + 
  {{{{{ (\tT_2)}}^3}\,{{{ t_6}}^2}}\over 6} - 
  {{2\,{ \tT_2}\,{ T_4}\,{{{ t_6}}^2}}\over 3} - 
  {{16\,{{{ t_6}}^3}}\over {27}} 
   - {{8\,{\phi^2}\,{ T_{10}}\,{ T_6}}\over 3} - 
  {{{{{ (\tT_2)}}^3}\,{ t_6}\,{ T_6}}\over {12}}\cr& + 
  {{{ \tT_2}\,{ T_4}\,{ t_6}\,{ T_6}}\over 3} + 
  {{8\,{{{ t_6}}^2}\,{ T_6}}\over 9} - 
  {{4\,{ t_6}\,{{{ T_6}}^2}}\over 9} + 
  {{2\,{{{ T_6}}^3}}\over {27}} - 
  2\,{\phi^2}\,{ \tT_2}\,{ t_6}\,{ T_8} - 
  {{{{{ (\tT_2)}}^2}\,{ t_6}\,{ T_8}}\over 6}\cr& + 
  {{2\,{ T_4}\,{ t_6}\,{ T_8}}\over 3} + 
  {{{{{ (\tT_2)}}^2}\,{ T_6}\,{ T_8}}\over {12}} - 
  {{{ T_4}\,{ T_6}\,{ T_8}}\over 3} + 
  {\phi^2}\,{{{ T_8}}^2}  }
}
where $\tT_2=T_2-\phi^2$.  The variables $T_n$
now refer to invariants of $SO(12)$.  $t_6$ is the product of six masses.

The $E_8$ representation splits into an adjoint and two fundamentals of
$E_7$.  The $E_7$ adjoint is made up of an $SO(12)$ adjoint, spinor 
and singlet, where the spinor and singlet are charged under the $SU(2)$.
The poles for the $SO(12)$ adjoint,
whose residues are  $\hij=m_i+m_j$ (plus sign permutations) 
divided by the usual factor
of $2\sqrt{2}\pi i$, satisfy
\eqn\xadjsev{\eqalign{
\tx_{ij,7}&=
-\left(\rho-w_4-\ha\ij\tW_2\right)^2+\hij^2\biggl(-\thi\rho\tW_2+\tW_2w_4
+{2\over3}W_6+\thi\ij(-\rho+2W_4+5w_4)\cr
&+\si(\ij)^2(\tW_2+4\phi^2)\biggr)
+\hij^4\left(-\thi\rho-\thi W_4-w_4-\qu(\ij)^2\right),
}}
where $W_n$ and $w_4$ refer to the $SO(8)$ invariants transverse to the 
$i$ and $j$ directions and where $\tW_2=W_2-\phi^2$.  The $\ty_{ij}$
for these poles are
\eqn\yadjsev{\eqalign{
\sqrt{-\ty_{ij}^2}=&\left(\rho-w_4-\ha\ij\tW_2\right)^3-
\qu\hij^2\left(\rho-w_4-\ha\ij\tW_2\right)\times\cr
&\qquad\times\Bigl(-2\rho\tW_2+6\tW_2w_4+4W_6+\ij(
-2\rho+4W_4+10w_4)+(\ij)^2(\tW_2+4\phi^2)\Bigr)\cr
&-\ei\hij^4\Bigl(4\rho^2+8w_4W_4+12w_4^2+4\tW_2W_6+2\tW_2^2w_4\cr
&\qquad\qquad +4\ij(-\rho\tW_2-
2\rho\phi^2+3\tW_2w_4+6\phi^2w_4+3W_6)\cr
&\qquad\qquad+2(\ij)^2(\rho^2+3w_4+W_4)-(\ij)^3(\tW_2+4\phi^2)\Bigr)\cr&
\ei\hij^6\Bigl(4w_4\tW_2+8w_4\phi^2+4W_6+2\ij(-\rho+3w_4)-(\ij)^3\Bigr)
-\qu\hij^8w_4.
}}

The $SO(12)$ singlet pole position, $x_7=\tx_7/h^2$, where $h=\phi$,
is given by
\eqn\xscsev{\eqalign{
\tx_7=&-\left(\rho-\ei(\tT_2)^2+\ha T_4\right)^2+\thi\phi^2(\rho\tT_2+T_6-2t_6)
}}
and $\ty_7$ is
\eqn\yscsev{\eqalign{
\sqrt{-\ty_7^2}=&\left(\rho-\ei(\tT_2)^2+\ha T_4\right)^3-
\phi^2\left(\rho-\ei(\tT_2)^2+\ha T_4\right)(\rho\tT_2+T_6-2t_6)
-\phi^4(\rho^2+\tT_2t_6-T_8).
}}
Because of the manner in which  $E_7$  was reached from
$E_8$, the $SO(12)$ spinor that is part of the $E_7$ adjoint has an odd 
number of minus signs.  The residue is $h_{sp}/(2\sqrt{2}\pi i)$, with
 $h_{sp}=(S_1-\phi)/2$, and where 
$S_1$ is the sum of $SO(12)$ masses with an odd number of minus signs.
It is straightforward to find the pole position from \xspin, with the result
\eqn\xspmsev{\eqalign{
x_{sp,7}=& -\trho^2+2\hsp\trho S_3 -{1\over3}\hsp^2(4\trho S_2+S_3^2+8S_5\phi)
+{2\over3}\hsp^3(\trho\phi+2S_4\phi-2S_5)\cr
&\qquad\qquad+{2\over3}\hsp^4(\trho+2S_4),
}}
where the $S_n$ are symmetric polynomials in the $m_i$ but with an odd 
number of $m_i$ replaced by $-m_i$ and the  term $\tilde \rho$ is $\tilde \rho=\rho
+S_4$  The value for $y$ at such a pole,
$\ty_{sp,7}$, is  given by
\eqn\yspmsev{\eqalign{
\sqrt{-\ty_{sp,7}^2}=&
\trho^3-3\hsp\trho^2S_3+2\hsp^2\trho(\trho S_2+S_3^2+2S_5\phi)
-\hsp^3\Bigl(\trho^2\phi+4S_3S_5\phi\cr
&\qquad+2\trho(S_2S_3+S_4\phi-2S_5)\Bigr)
-2\hsp^4\Bigl(\trho^2-\trho S_3\phi+2S_3S_5-2S_2S_5\phi-2S_6\phi^2\Bigr)\cr
&\qquad+2\hsp^5(\trho Q_3+2S_2S_5+2S_6\phi-2S_5\phi^2)-8\hsp^6S_5\phi
-4\hsp^7S_5.
}}

The $E_7$ fundamental has  an $SO(12)$ vector and the other spinor.
The vector poles are at
\eqn\xvecsev{\eqalign{
x_{i,7}=&
  \rho\,\left( {{5\,{{{ m_i}}^2}}\over 3} + 2\,{ m_i}\,\phi + 
     {{{\phi^2}}\over 3} - {{{ W_2}}\over 3} \right)
-{{{{{ m_i}}^6}}\over 4} + 
  {{2\,{ m_i}\,{ w_5}}\over 3} - {{{ m_i}}^5}\,\phi + 
  2\,{ w_5}\,\phi - {{3\,{{{ m_i}}^4}\,{\phi^2}}\over 2} - 
  {{{ m_i}}^3}\,{\phi^3} \cr&
- {{{{{ m_i}}^2}\,{\phi^4}}\over 4}  + 
  {{{{{ m_i}}^4}\,{ W_2}}\over 2} + 
  {{{ m_i}}^3}\,\phi\,{ W_2} + 
  {{{{{ m_i}}^2}\,{\phi^2}\,{ W_2}}\over 2} - 
  {{{{{ m_i}}^2}\,{{{ W_2}}^2}}\over 4} + 
  {{2\,{{{ m_i}}^2}\,{ W_4}}\over 3} - {{{ W_6}}\over 3}
}}
with
\eqn\yvecsev{\eqalign{
\sqrt{-y_{i,7}^2}=&(\phi+2z_i)\rho^2+\left(2w_5+W_4z_i-{1\over4}\left((\phi+z_i)^2-W_2\right)^2-z_i^2(\phi+z_i)\left((\phi+z_i)^2-W_2\right)\right)\rho\cr
&+{1\over8}z_i^3\left((\phi+z_i)^2-W_2\right)^3
-z_i^3W_4\left((\phi+z_i)^2-W_2\right)\cr
\qquad\qquad&-w_5\left(((\phi+z_i)^2-R_2)^2
+2z_i^2(\phi+z_i)\right)+W_6\phi z_i^2-W_8\phi}
}

The spinor poles are at
\eqn\xspsev{\eqalign{
x_{sp',7}=& 
  \rho\,\left( {{{{{ s_1}}^2}}\over 6} + 
     {{2\,{ s_2}}\over 3} - {{{\phi^2}}\over 6} \right)
-{{{{{ s_3}}^2}}\over 3} - 
  {{{{{ s_1}}^2}\,{ s_4}}\over 2} + 
  {{2\,{ s_2}\,{ s_4}}\over 3} + 
  {{4\,{ s_1}\,{ s_5}}\over 3} - {{8\,{ s_6}}\over 3} + 
  {{{ s_4}\,{\phi^2}}\over 2} 
}}
where $s_n$ are symmetric polynomials with an even number of
minus signs.  The $y$ values at the poles are
\eqn\yspsev{\eqalign{
\sqrt{-y_{sp,7}^2}=&{\rho^2}\,{ s_1} + 
  \rho\,\left( + 2\,{ s_5} + {{{ s_3}\,{\phi^2}}\over 2} - 
{{\left( {{{ s_1}}^2}\,{ s_3} \right) }\over 
       2}\right)
 + {{{{{ s_1}}^2}\,{ s_3}\,{ s_4}}\over 
    2} - { s_1}\,{{{ s_4}}^2} + 
  {{{{{ s_1}}^4}\,{ s_5}}\over 4}\cr& - 
  {{{ s_1}}^2}\,{ s_2}\,{ s_5} + 2\,{ s_4}\,{ s_5} - 
  {{{ s_1}}^3}\,{ s_6} + 4\,{ s_1}\,{ s_2}\,{ s_6} - 
  4\,{ s_3}\,{ s_6} - 
  {{{ s_3}\,{ s_4}\,{\phi^2}}\over 2} - 
  {{{{{ s_1}}^2}\,{ s_5}\,{\phi^2}}\over 2}\cr& + 
  { s_2}\,{ s_5}\,{\phi^2} + { s_1}\,{ s_6}\,{\phi^2} + 
  {{{ s_5}\,{\phi^4}}\over 4}
}}
Notice that in \yvecsev\  and \yspsev, the coefficient in front of the $\rho^2$
term is twice $h_\al$.

\appendix{C}{$E_6$ Results}

The $E_6$ curve is reached by letting $m_6=\Lambda$ and $\phi=\Lambda-6\la$,
while scaling $x\to x\Lambda^2$, $y\to y\Lambda^3$ and $\rho\to \rho\Lambda$.
The resulting expressions for $f$ and $g$ are
\eqn\feqEsix{\eqalign{
f_{E_6}=&
{\rho^2}\,\left( 12\,{\lambda^2} + { T_2} \right)  + 
  \rho\,\left( 8\,\lambda\,{ T_4} + 8\,{ t_5} \right) \cr& + 
  36\,{\lambda^2}\,{ T_6} - { T_2}\,{ T_6} + 4\,{ T_8} + 
  t_5{{{{{ T_4}}^2}}\over 3}
 - 432\,{\lambda^3}\,{ } + 
  12\,\lambda\,{ T_2}\,{ t_5}
}}
\eqn\geqEsix{\eqalign{
g_{E_6}=&
{\rho^4} + {\rho^3}\,\left( -16\,{\lambda^3} + 4\,\lambda\,{ T_2} \right) + 
  {\rho^2}\,\left( 20\,{\lambda^2}\,{ T_4} + 
     {{{ T_2}\,{ T_4}}\over 3} - 40\,\lambda\,{ t_5} - 
     2\,{ T_6} \right) \cr&
 + \rho\,\biggl( {{8\,\lambda\,{{{ T_4}}^2}}\over 3} + 
     864\,{\lambda^4}\,{ t_5} - 96\,{\lambda^2}\,{ T_2}\,{ t_5} + 
     2\,{{{ T_2}}^2}\,{ t_5} - 
     {{16\,{ T_4}\,{ t_5}}\over 3} + 
     144\,{\lambda^3}\,{ T_6}\cr&
 - 4\,\lambda\,{ T_2}\,{ T_6} - 
     32\,\lambda\,{ T_8} \biggr) \cr&
 + 
  {{2\,{{{ T_4}}^3}}\over {27}} - 
  144\,{\lambda^3}\,{ T_4}\,{ t_5} + 
  4\,\lambda\,{ T_2}\,{ T_4}\,{ t_5} + 
  144\,{\lambda^2}\,{{{ t_5}}^2}
   + 12\,{\lambda^2}\,{ T_4}\,{ T_6} - 
  {{{ T_2}\,{ T_4}\,{ T_6}}\over 3}\cr& - 
  24\,\lambda\,{ t_5}\,{ T_6} + {{{ T_6}}^2} + 
  1296\,{\lambda^4}\,{ T_8} - 72\,{\lambda^2}\,{ T_2}\,{ T_8} + 
  {{{ T_2}}^2}\,{ T_8} - 
  {{8\,{ T_4}\,{ T_8}}\over 3}
}}

The $E_6$ adjoint is made up of the $SO(10)$ adjoint and two spinors.
The spinors are charged under the $U(1)$, with the separate spinor
representations having opposite charges.
The $SO(10)$ adjoint poles are  positioned at $\tx_{ij,6}=x_{ij,6}\hij$,
where $\hij=m_i+m_j$ and
\eqn\xadjsix{\eqalign{
\tx_{ij,6}&= -\left(\rho-6\ij\la-w_3\right)^2+
{2\over3}\hij^2\Bigl(-6\la(\rho-3w_3)+W_4+\ij W_2+(\ij)^2\Bigr)\cr
&\qquad\qquad-\thi\hij^4W_2,
}}
where $W_n$ and $w_3$ are the $SO(6)$ invariants orthogonal to $i$ and
$j$.
At these points, $\ty_{ij,6}$ is given by
\eqn\yadjsix{\eqalign{
\sqrt{-\ty_{ij,6}^2}&= \left(\rho-6\ij\la-w_3\right)^3+
\hij^2\left(\rho-6\ij\la-w_3\right)\times\cr
&\qquad\qquad\qquad\times\Bigl(6\la(\rho-3w_3)-W_4-\ij W_2-(\ij)^2\Bigr)\cr
&\qquad\qquad
-\hij^4\Bigl(w_3(W_2+36\la^2)+6\la W_4-\ij(\rho-3w_3)\Bigr)+\hij^6w_3
}}

The SO(10) spinors in the $E_6$ adjoint have poles at $x\hsp^2=\tx_{sp,6}$,
where $\hsp=S_1/2+3\la$ and
\eqn\xspmsix{\eqalign{
x_{sp,6}=&-\trho^2+2\hsp\trho S_2+\hsp^2\left(8\trho\la-6\la S_3-\thi S_2^2-
{8\over3}S_4\right)-{2\over3}\hsp^3(3\trho-2S_3)}}
where the $S_n$ are the symmetric polynomials over the five $m_i$ of $SO(10)$
but with an odd number of $m_i$ replaced with $-m_i$.  The variable 
$\trho$ is $\trho=\rho+S_3$.
At these poles, $\ty_{sp}=y\hsp^3$ satisfies
\eqn\yspmsix{\eqalign{
\sqrt{-\ty_{sp,6}^2}=&\trho^3-3\hsp S_2\trho^2-
2\hsp^2\trho(6\trho\la-S_2^2-2S_4)\cr
&+\hsp^3(\trho(3\trho-2S_3)+12\trho\la S_2-4S_2S_4)+2\hsp^4
(-\trho S_2-12\la S_4+2S_5).
}}

The $E_6$ fundamental poles were given in \MNIII, which we repeat here for
convenience.  The $SO(10)$ vector pole positions and values for $y$ are
\eqn\xvsix{\eqalign{
x_{i,6}=& 
 \rho \left( -4\,{ \lambda} + 2\,{ m_i} \right)
-36\,{{{ \lambda}}^2}\,{{{ m_i}}^2} + 
  12\,{ \lambda}\,{{{ m_i}}^3} - {{{ m_i}}^4}  + 
  2\,{ w_4} + {{2\,{{{ m_i}}^2}\,{ W_2}}\over 3} - 
  {{{ W_4}}\over 3}
}}
\eqn\yvsix{\eqalign{
\sqrt{-y_{i,6}^2}=& {\rho^2}+ \rho\,
   \left( -36\,{\lambda^2}\,{ m_i} + 24\,\lambda\,{{{ m_i}}^2} - 
     3\,{{{ m_i}}^3} + { m_i}\,{ W_2} \right) 
-216\,{\lambda^3}\,{{{ m_i}}^3} + 108\,{\lambda^2}\,{{{ m_i}}^4}\cr&
 -  18\,\lambda\,{{{ m_i}}^5}
 + {{{ m_i}}^6} - 36\,{\lambda^2}\,{ w_4} + 12\,\lambda\,{ m_i}\,{ w_4} - 
  3\,{{{ m_i}}^2}\,{ w_4} + 
  6\,\lambda\,{{{ m_i}}^3}\,{ W_2}  \cr
&\qquad\qquad-{{{ m_i}}^4}\,{ W_2} + 
  { w_4}\,{ W_2}  + 
  {{{ m_i}}^2}\,{ W_4} - { W_6}
}}
where $W_n$ and $w_4$ are the $SO(8)$ invariants orthogonal to $i$.  The
residues are $h_i=m_i-2\la$ divided by $2\sqrt{2}\pi i$.
The spinor poles satisfy
\eqn\xspsix{\eqalign{
x_{sp',6}=&
\rho\,\left( 2\,\lambda + { s_1} \right)  - 
  {{{{{ s_2}}^2}}\over 3} - 6\,\lambda\,{ s_3} - 
  {{{ s_1}\,{ s_3}}\over 3} + {{4\,{ s_4}}\over 3}
}}
\eqn\yspsix{\eqalign{
\sqrt{-y_{sp',6}^2}=&
{\rho^2} + \rho\,\left( -6\,\lambda\,{ s_2} - { s_1}\,{ s_2} \right) 
   + 6\,\lambda\,{ s_2}\,{ s_3} + { s_1}\,{ s_2}\,{ s_3} - 
  {{{ s_3}}^2} + 36\,{\lambda^2}\,{ s_4}\cr
&\qquad\qquad - 
  {{{ s_1}}^2}\,{ s_4} - 12\,\lambda\,{ s_5} + 
  2\,{ s_1}\,{ s_5}
}}
where $s_n$ are symmetric polynomials with an {\it even} number of
$-$ signs.  The residues are $(s_1/2+\la)/(2\sqrt{2}\pi i)$.
Finally, the $SO(10)$ singlet, with residue $4\la/(2\sqrt{2}\pi i)$ 
has poles at
\eqn\xsingsix{\eqalign{x_{s,6}=
  8\,\lambda\,\rho-324\,{\lambda^4} + 18\,{\lambda^2}\,{ T_2} - 
  {{{{{ T_2}}^2}}\over 4} + {{2\,{ T_4}}\over 3}
}}
with
\eqn\ysingsix{\eqalign{
\sqrt{-y_{s,6}^2}=&
 {\rho^2} +5832\,{\lambda^6} - 486\,{\lambda^4}\,{ T_2} + 
  {{27\,{\lambda^2}\,{{{ T_2}}^2}}\over 2} - 
  {{{{{ T_2}}^3}}\over 8} + 
  \rho\,\left( -216\,{\lambda^3} + 6\,\lambda\,{ T_2} \right) \cr
&\qquad\qquad
 -  18\,{\lambda^2}\,{ T_4} + {{{ T_2}\,{ T_4}}\over 2} + 
  12\,\lambda\,{ T_5} - { T_6}
}}

\appendix{D}{$D_4$ Results}

For the $D_4$ case, the appropriate scaling is $\la=-c_1\Lambda/6$,
 $m_5=-c_2\Lambda$, $\rho=u\Lambda$, $x\to x\Lambda^2$, and
 $y\to y\Lambda^3$, where $c_1$ and $c_2$ are defined in \SWII.  Keeping
the leading order in $\Lambda$, the $f$ and $g$ terms reduce to 
\eqn\fdfour{\eqalign{
f_{D_4}= &
  \left( {{{{{ c_1}}^2}}\over 3} + {{{ c_2}}^2} \right) \,
   {u^2}- 
  {{4\,{ c_1}\,{{{ c_2}}^2}\,{ T_2}\,u}\over 3}
-2\,{{{ c_1}}^3}\,{ c_2}\,{ t_4} + 
  2\,{ c_1}\,{{{ c_2}}^3}\,{ t_4} + 
  {{{{{ c_2}}^4}\,{{{ T_2}}^2}}\over 3}\cr& + 
  {{{ c_1}}^2}\,{{{ c_2}}^2}\,{ T_4} - 
  {{{ c_2}}^4}\,{ T_4}
}}
and
\eqn\gdfour{\eqalign{ 
g_{D_4}=& 
  \left( {{2\,{{{ c_1}}^3}}\over {27}} - 
     {{2\,{ c_1}\,{{{ c_2}}^2}}\over 3} \right) \,{u^3} + 
\left( {{5\,{{{ c_1}}^2}\,{{{ c_2}}^2}\,
	 { T_2}}\over 9} + 
     {{{{{ c_2}}^4}\,{ T_2}}\over 3} \right) \,{u^2}\cr& + 
  \left( {{-2\,{{{ c_1}}^4}\,{ c_2}\,{ t_4}}\over 3} + 
     {{8\,{{{ c_1}}^2}\,{{{ c_2}}^3}\,{ t_4}}\over 3} - 
     2\,{{{ c_2}}^5}\,{ t_4} - 
     {{4\,{ c_1}\,{{{ c_2}}^4}\,{{{ T_2}}^2}}\over 9} - 
     {{2\,{{{ c_1}}^3}\,{{{ c_2}}^2}\,{ T_4}}\over 3} + 
     {{2\,{ c_1}\,{{{ c_2}}^4}\,{ T_4}}\over 3} \right) 
    \,u\cr&
-{{2\,{{{ c_1}}^3}\,{{{ c_2}}^3}\,{ t_4}\,{ T_2}}\over 
    3} + {{2\,{ c_1}\,{{{ c_2}}^5}\,{ t_4}\,
      { T_2}}\over 3} + 
  {{2\,{{{ c_2}}^6}\,{{{ T_2}}^3}}\over {27}} + 
  {{{{{ c_1}}^2}\,{{{ c_2}}^4}\,{ T_2}\,{ T_4}}\over 
    3} - {{{{{ c_2}}^6}\,{ T_2}\,{ T_4}}\over 3}\cr& + 
  {{{ c_1}}^4}\,{{{ c_2}}^2}\,{ T_6} - 
  2\,{{{ c_1}}^2}\,{{{ c_2}}^4}\,{ T_6} + 
  {{{ c_2}}^6}\,{ T_6}.
}}
The $T_n$ and $t_4$ are $SO(8)$ invariants.
The pole positions for the $SO(8)$ adjoint, with $\hij=\pm m_i\pm m_j$
satisfy
\eqn\xadjfour{\eqalign{
\tx_{ij,4}=& -(u+c_2w_2+c_1\ij)^2+{2\over3}\hij^2(c_1u+c_2^2W_2+3c_1c_2W_2+
c_2^2\ij)-\thi c_2^2\hij^4,
}}
where $W_2$ and $w_2$ refer to the $SO(4)$ invariants transverse to the
$i$ and $j$ directions.  The values of $\sqrt{-\ty_{ij,4}^2}$ 
at these poles are
\eqn\yadjfour{\eqalign{
\sqrt{-\ty_{ij,4}^2}=&(u+c_2w_2+c_1\ij)^3-\hij^2(u+c_2w_2+c_1\ij)
(c_1u+c_2^2W_2+3c_1c_2W_2+c_2^2\ij)\cr
&\qquad\qquad+\hij^4c_2(c_1c_2W_2+(c_1^2+c_2^2)w_2).}}

\listrefs  
\end
\end